\documentclass[aps,twocolumn,groupedaddress,floatfix,longbibliography,superscriptaddress]{revtex4-1}
\usepackage{amsmath}
\usepackage{amssymb}
\usepackage{amsfonts}
\usepackage{graphicx}
\usepackage{bbold}
\usepackage{bm}
\usepackage{bm}
\usepackage{times,float}
\usepackage{graphicx}
\usepackage[usenames,dvipsnames,svgnames]{xcolor}
\usepackage{hyperref}
\hypersetup{colorlinks=true, linkcolor=Blue, citecolor=Green,urlcolor=Blue}
\usepackage{braket}
\usepackage{accents}

\newcommand{\be}{\begin{equation}}
\newcommand{\ee}{\end{equation}}
\newcommand{\bea}{\begin{eqnarray}}
\newcommand{\eea}{\end{eqnarray}}
\newcommand{\beas}{\begin{eqnarray*}}
\newcommand{\eeas}{\end{eqnarray*}}
\newcommand{\nn}{\nonumber}

\newcommand{\hrho}{\hat{\rho}}
\newcommand{\hvrr}{\hat{\varrho}}
\newcommand{\hH}{\hat{\mathcal{H}}}
\newcommand{\ah}{\hat{a}}
\newcommand{\ad}{\hat{a}^\dagger}
\newcommand{\bs}{\beta^\star}

\begin{document}
\title{Entropy exchange and thermal fluctuations in the Jaynes-Cummings model}
\author{Jorge David Casta\~no-Yepes}\email{ jorgecastanoy@gmail.com}
\affiliation{Facultad de Ciencias - CUICBAS, Universidad de Colima, Bernal D\'iaz del Castillo No. 340, Col. Villas San Sebasti\'an, 28045 Colima, Mexico.}
\begin{abstract}
{ The time-dependence of the quantum entropy for a two-level atom interacting with a single-cavity mode is computed using the Jaynes-Cummings model, when the initial state of the radiation field is prepared in a thermal state with temperature fluctuations. In order to describe the out-of-equilibrium situation, the Super Statistics approximation is implemented so that the gamma and the multi-level distribution functions are used to introduce the inverse temperature fluctuations. In the case of the gamma distribution, paralleling the Tsallis non-additive formalism, the entropy for the system is computed with the $q$-logarithm prescription, and the impact of the initial state of the atom is also taken into account. The results show that, in the first distribution, the $q$-parameter (related to the thermal fluctuations) modifies the partial entropies appreciably. In contrast, the way the inverse temperatures are distributed in the second one may lead to changes in the entropy functions.}
\end{abstract}
\maketitle

\section{Introduction}\label{Sec:Intro}
In the foundations and application of quantum mechanics, entanglement is an interesting feature that gives an understanding of non-local correlations, with several applications in quantum information processing and quantum computation~\cite{PhysRevLett.67.661,nielsen2002quantum,kjaergaard2020superconducting,steane1998quantum}. Regarding the description of the entanglement, the entropy is a key concept~\cite{PhysRevA.102.052407,PhysRevLett.103.140401}, and its study covers a wide range of research topics like coherent information~\cite{PhysRevA.76.014306,PhysRevA.101.062101}, long-range interactions~\cite{PhysRevA.97.062301,PhysRevA.92.042334}, quantum-phase transitions and thermodynamics~\cite{shafieinejad2020entanglement,PhysRevA.66.032110,PhysRevLett.90.227902}, among others~\cite{PhysRevA.97.022326,PhysRevLett.80.2245,gigena2013generalized,bengtson2017role}. 

In general terms, the entropy is related to the time evolution of the quantum state, so that if the dynamics is divided into a system in contact with a thermal bath, the coupling between them leads to entropy exchange. The latter requires the specification of the initial quantum configuration of the system and bath. The future behavior can be computed from different methods, such as unitary time evolution or the master equation formalism. It is common to factorize the initial density operator as { $\hrho_a\otimes\hrho_b$, where $\hrho_b$ is the thermal state associated with the reservoir, which is considered much larger than the system (described by $\hrho_a$)}, and therefore, it cannot be affected by the system influence. Nevertheless, what happens if the bath has thermal fluctuations? Even if the latter is bigger than the system, an out-of-equilibrium  situation should be present in the initial density operator $\hrho_b$, and as time passes, that information may modify the own dynamics of $\hrho_a$.

The discussion above highlights the need to accomplish a theory for non-equilibrium thermodynamics, which has been approximated in several ways~\cite{muschik1990aspects,lebon2008understanding,de2013non}. One of the attempts to describe such a situation is the so-called Super-Statistics (SS), developed by Beck and Cohen to study processes where some intensive parameter fluctuates~\cite{beck2003superstatistics,beck2004superstatistics,beck2009recent}. In this framework, if it is assumed that the system passes for stages that are close to the thermal equilibrium, a generalized or modified Boltzmann factor can be postulated. The latter is found as an average over the local equilibrium dynamics with the help of a probability distribution function. The SS idea has been applied in different research areas, such as cosmic rays~\cite{beck2004generalized}, high-energy physics~\cite{PhysRevD.98.114002,PhysRevD.91.114027,beck2009superstatistics}, turbulence models~\cite{PhysRevLett.91.084503,PhysRevE.72.026304}, and other topics~\cite{PhysRevE.88.062146,PhysRevE.95.042111,PhysRevE.72.056133,gravanis2021blackbody}.

To apply the SS ideas to elucidate the consequences of thermal fluctuation in the entropy exchange of two quantum systems, the Jaynes-Cummings model which describes a two-level atom coupled with a single-mode quantized field, is used~\cite{jaynes1963comparison}. In the rotating wave approximation, this model is a prototype to study the partial entropy exchange and entanglement with analytical treatment~\cite{PhysRevA.44.6023,PhysRevA.45.8190,PhysRevA.71.063821,guo2011entropy}. The fluctuations in the temperature of the cavity mode are introduced by writing the initial quantum state $\hrho_b$ in terms of the modified Boltzmann factor provided by SS. Two density distribution functions are chosen: a gamma or $\chi^2$-distribution function, closely related to the non-additive Tsallis formalism, and a multi-level distribution function. For the first one, by imposing constraints over the distribution free parameters, and demanding a formalism that preserves the Legendre structure of thermodynamics, the well-known Tsallis non-additive entropy is found. For the second distribution choice, aleatory inverse temperatures are simulated to understand the importance of the fluctuations.  { Then, in this paper, the effects of non-equilibrium dynamics are encoded in the variation of the parameters corresponding to each distribution function}. Also, the contributions of the initial state $\hrho_a$, are studied as a function of the excited and ground state superposition. The results show that the thermal fluctuations on the initial cavity mode configuration deeply impact the time-dependent entropy exchange.

The paper is organized as follows: Sec.~\ref{Sec:SS} covers the formal aspects of SS, and introduce the distribution functions commented above. Sec.~\ref{Sec:JCmodel} presents the Jaynes-Cummings model (JCM) and the general form of the entropy exchange function. In Sec.~\ref{Sec:EntropyExchange}, the initial density operators for the cavity field are given in an explicit form, by following the modified Boltzmann factors associated with the selected distribution functions. Sec.~\ref{Sec:Results} presents the results and their discussion, and finally, Sec.~\ref{Sec:Concl} provides corresponding conclusions. Additionally, a short review of the Tsallis non-additive formalism is given in Appendix~\ref{Sec:SSandTsallis}. In Appendix~\ref{Sec:Naverage}, the calculation of the average photon number in the non-additive case is performed.

\section{Super-Statistics}\label{Sec:SS}
Super statistics is a novel approximation to the non-equilibrium thermodynamics, where the fluctuations of some intensive parameter $\widetilde{\beta}$ (inverse temperature, chemical potential, noise, mass variation) are introduced through a probability density function $f(\widetilde{\beta})$. In its general form, if the system admits states near to the equilibrium, described by Gibbs-Boltzmann probabilities, the SS framework is constructed from a modified Boltzmann factor $\hat{B}$, given by
\bea
\hat{B}=\int_0^\infty d\tilde{\beta} f(\tilde{\beta}) e^{-\tilde{\beta}\hat{H}},
\label{Bdef}
\eea
where $\hat{H}$ is the Hamiltonian of the system. Then, SS is a superposition of two statistics: one, referred to the local equilibrium with Boltzmann factor $e^{-\tilde{\beta}\hat{H}}$ and the other given by the probability density distribution $f(\tilde{\beta})$. It is necessary to point out that SS  does not constitute a statistical mechanics theory because Eq.~(\ref{Bdef}) is an \textit{ansatz}. Nevertheless, entropic functionals and constraints in the thermodynamic potentials can be found in order to resemble the modified Boltzmann factor~\cite{PhysRevE.67.026106}.

In general, $f(\tilde{\beta})$ is arbitrary and its choice responds to the particular scenario in which SS is applied. In this paper, the so-called gamma and multi-level density distribution functions were used to obtain analytical results, given by
\bea
f(\tilde{\beta})=\frac{1}{b\Gamma(c)}\left(\frac{\tilde{\beta}}{c}\right)^{c-1}e^{-\tilde{\beta}/b},
\label{chisquared}
\eea
and
\bea
f(\tilde{\beta})=\frac{1}{N}\sum_{k=1}^N\delta(\widetilde{\beta}-\beta_k).
\label{multileveldist}
\eea

In Eq.~(\ref{chisquared}), $\Gamma(x)$ is the Euler-gamma function, and $c$ and $b$ are free parameters. On the other hand in Eq.~(\ref{multileveldist}) $\delta(x)$ is the Dirac-delta function and the sum runs over $N$-values of discrete intensive parameters.

To connect with a well-established thermodynamics framework, i.e., the Tsallis non-additive formalism~\cite{tsallis1998role}, the free parameters of the gamma distribution are chosen as $bc=\beta$ and $c=1/(q-1)$, where $\beta$ is identified with the average temperature of the system:
\bea
\langle\widetilde{\beta}\rangle=\int_0^\infty d\tilde{\beta} f(\widetilde{\beta})\widetilde{\beta}=\beta.
\eea

Then, the modified Boltzmann factor reads:
\bea
\hat{B}=e_q^{-\beta\hat{H}}
\label{B1eq}
\eea
where the $q$-exponential is defined as:
\bea
e_q^x\equiv[1+(1-q)x]^{1/(1-q)},
\eea
and the Boltzmann factor $e^{-\beta\hat{H}}$ is recovered in the limit $q\to1$.

The form of parameters $a$ and $b$ is chosen to give an account of a genuinely physical temperature from the particular statistical weight factor of Eq.~(\ref{B1eq}). The latter has foundations in preserving the Legendre structure, and it was used in previous works to emphasize the differences between thermal-dependent functions with and without the Tsallis energy constraints~\cite{castano2020super,castano2020comments,castano2021super}. Therefore, in this work, the application of the gamma-distribution needs to be understood as if the thermal fluctuations lead to  non-additive thermodynamics. For completeness, a brief review of the Tsallis formalism is given in Appendix~\ref{Sec:SSandTsallis}.

For the multi-level density distribution function, the modified Boltzmann factor can be obtained:
\bea
\hat{B}=\frac{1}{N}\sum_{k=1}^Ne^{-\beta_k\hat{H}},
\label{BNlevel}
\eea
so that the usual statistical mechanics is obtained when all the $\beta_k$ are the same.

Note that the trace of Eq.~(\ref{BNlevel}) can define a super-partition function $Z_N$ which has the same mathematical structure as the {\it Boltzmannian} one:
\bea
Z_N=\text{Tr}\left[\hat{B}\right]=\sum_{n,k} e^{\beta\widetilde{E}_{nk}},
\eea
where if $E_n$ are the eigenvalues of the Hamiltonian, the spectrum can be redefined as:
\bea
\widetilde{E}_{nk}=\frac{\beta_k}{\beta}E_n,
\eea
and a physical temperature $\beta$ of the whole system is defined.

In the following sections, both SS formalism are applied to elucidate the impact on the entropy exchange by thermal fluctuations in the cavity mode initial state, when the full system is modeled with the Jaynes-Cummings model.
\section{Jaynes-Cummings model and density operators}\label{Sec:JCmodel}
The JCM for a two-level atom with a single-mode radiation field is given by:
\bea
\hH=\frac{\omega_0}{2}\hat{\sigma}_z+\omega\ad\ah+\frac{\lambda}{2}\left(\ad\hat{\sigma}_{-}+\ah\hat{\sigma}_{+}\right).
\label{Hamiltonian}
\eea

{ The eigenstates and eigenvalues (including the so-called {\it uncoupled state}) of the Hamiltonian are:
\bea
\ket{\psi_0}&=&\ket{g,0}\nn\\
\ket{\psi_n^\pm}&=&\frac{1}{\sqrt{1+\Omega_\pm^2}}\left(\ket{e,n}+\Omega_\pm\ket{g,n+1}\right),
\eea
and
\bea
E_0&=&-\frac{\omega_0}{2},\nn\\
E_n^\pm&=&\omega\left(n+\frac{1}{2}\right)\pm\frac{\delta_n}{2},
\eea
where $\Delta=\omega_0-\omega$ is the detuning frequency and
\bea
\delta_n^2&=&\Delta^2+\lambda^2(n+1),\nn\\
\Omega_\pm&=&\frac{\Delta\pm\delta_n}{\lambda\sqrt{n+1}}.
\eea
}

If at the time $t=0$ the system is prepared in a state given by its density operator $\hat{\rho}(0)=\hat{\rho}_a(0)\otimes\hat{\rho}_b(0)$, where $\hat{\rho}_a(0)$ and $\hat{\rho}_b(0)$ are the density operators of the atom and the cavity field, respectively, the unitary time-evolution of $\hat{\rho}$ is given by
\bea
\hat{\rho}(t)&=&\hat{U}(t)\hrho(0)\hat{U}^\dagger(t),
\eea
where { $\hat{U}(t)$ is the evolution operator, which in the basis $\left\{\ket{\psi_0},\ket{\psi_n^\pm}\right\}$ takes the form:
\bea
\hat{U}(t)&=&e^{i\omega_0t/2}\ket{g,0}\bra{0,g}\nn\\
&+&\sum_{n=0}^{\infty}\left[e^{-i E_n^+t}\ket{\psi_n^+}\bra{\psi_n^+}+e^{-i E_n^-t}\ket{\psi_n^-}\bra{\psi_n^-}\right].\nn\\
\eea}

Therefore, if the cavity mode is prepared in the thermal state:
\bea
\hrho_b(0)=\sum_{n=0}^\infty p_n\ket{n}\bra{n},
\eea
and the atom is initially in the mixed state:
\bea
\hrho_a(0)=\epsilon\ket{e}\bra{e}+(1-\epsilon)\ket{g}\bra{g},\quad 0\leq\epsilon\leq1,
\eea
the density operator of the full system in the { {\it dressed}-state representation} $\left\{\ket{g,0},\ket{e,n},\ket{g,n+1}\right\}$ is{~\cite{guo2011entropy}}:
\bea
\hat{\rho}(t)&=&p_0(1-\epsilon)\ket{g,0}\bra{g,0}\\
&+&\sum_{n=0}^\infty\Big[\mathcal{A}_n(t)\ket{e,n}\bra{e,n}+\mathcal{B}_n(t)\ket{e,n}\bra{g,n+1}\nn\\
&+&\mathcal{B}_n^*(t)\ket{g,n+1}\bra{e,n}+\mathcal{C}_n\ket{g,n+1}\bra{g,n+1}\Big],\nn
\eea
where
\begin{subequations}
\bea
\mathcal{A}_n(t)&=&\frac{\epsilon\, p_n+(1-\epsilon)\Omega_+^2p_{n+1}}{(1+\Omega_+^2)^2}+\frac{\epsilon\, p_n+(1-\epsilon)\Omega_-^2p_{n+1}}{(1+\Omega_-^2)^2}\nn\\
&+&2\frac{\epsilon\,p_n+(1-\epsilon)\Omega_+\Omega_-p_{n+1}}{\left(1+\Omega_+^2\right)\left(1+\Omega_-^2\right)}\cos\left(\delta_n t\right)
\eea
\bea
\mathcal{B}_n(t)&=&\frac{\Omega_+\left[\epsilon\, p_n+(1-\epsilon)\Omega_+^2p_{n+1}\right]}{(1+\Omega_+^2)^2}\nn\\
&+&\frac{\Omega_-\left[\epsilon\, p_n+(1-\epsilon)\Omega_-^2p_{n+1}\right]}{(1+\Omega_-^2)^2}\\
&+&\frac{\epsilon\,p_n+(1-\epsilon)\Omega_+\Omega_-p_{n+1}}{\left(1+\Omega_+^2\right)\left(1+\Omega_-^2\right)}\left(\Omega_+e^{i\delta t}+\Omega_-e^{-i\delta t}\right),\nn
\eea
and
\bea
\mathcal{C}_n(t)&=&\frac{\Omega_+^2\left[\epsilon\, p_n+(1-\epsilon)\Omega_+^2p_{n+1}\right]}{(1+\Omega_+^2)^2}\nn\\
&+&\frac{\Omega_-^2\left[\epsilon\, p_n+(1-\epsilon)\Omega_-^2p_{n+1}\right]}{(1+\Omega_-^2)^2}\\
&+&\frac{2\Omega_+\Omega_-\left[\epsilon\,p_n+(1-\epsilon)\Omega_+\Omega_-p_{n+1}\right]}{\left(1+\Omega_+^2\right)\left(1+\Omega_-^2\right)}\cos\left(\delta_nt\right).\nn
\eea
\end{subequations}

From the former, the reduced density operators for the atom and cavity are:
\bea
\hrho_a(t)&\equiv&\text{Tr}_b\hrho(t)\\
&=&\left[\sum_{n=0}^\infty\mathcal{A}_n(t)\right]\ket{e}\bra{e}\nn\\
&+&\left[p_0(1-\epsilon)+\sum_{n=1}^\infty\mathcal{C}_{n-1}(t)\right]\ket{g}\bra{g},\nn
\eea
and
\bea
\hrho_b(t)&\equiv&\text{Tr}_a\hrho(t)\\
&=&\left[p_0(1-\epsilon)+\mathcal{A}_0\right]\ket{0}\bra{0}\nn\\
&+&\sum_{n=1}^\infty\left[\mathcal{A}_n(t)+\mathcal{C}_{n-1}(t)\right]\ket{n}\bra{n}\nn.
\eea

With the last expressions, the partial entropy for each subsystem can be computed.

\section{Entropy exchange and temperature fluctuations}\label{Sec:EntropyExchange}
Now, let me assume that the cavity field is prepared in a thermal state with fluctuations in its temperature. In order to model such a situation, and in according to Sec.~\ref{Sec:SS}, two density operators are considered, depending on which temperature density distribution function for fluctuations is selected:
\begin{enumerate}
    \item For gamma-distribution function:
    \bea
    \hat{\rho}_b(0)=\frac{1}{Z_b}\exp_q\left(-\beta\frac{\hH_b-U_b}{\text{Tr}\left[\hat{\rho}_b^q(0)\right]}\right),
    \label{rhob1}
    \eea
    where $\hH_b=\omega\ad\ah$,
    \bea
    Z_b=\text{Tr}\left[\exp_q\left(-\beta\frac{\hH_b-U_b}{\text{Tr}\left[\hat{\rho}_b^q(0)\right]}\right)\right],
    \label{Z}
    \eea
    and
    \bea
    U_b=\frac{\text{Tr}\left[\hat{\rho}_b^q(0)\hH_b\right]}{\text{Tr}\left[\hat{\rho}_b^q(0)\right]}.
    \label{U}
    \eea

    \item For an $N$-level distribution function:
    \bea
    \hat{\rho}_b(0)=\frac{1}{Z_N}\sum_{k=1}^N e^{-\beta_k\omega\ad\ah},
    \label{rhoNlevel}
    \eea
    where
    \bea
    Z_N=\sum_{k=1}^{N}e^{\beta_k\omega}\bar{n}(\omega_k,\beta_1)=N+\sum_{k=1}^{N}\bar{n}(\omega_k,\beta_1)
    \eea
    and
    \bea
    \bar{n}(\omega,\beta)=\frac{1}{e^{\beta\omega}-1}.
    \eea
\end{enumerate}

Equations~(\ref{rhob1})-(\ref{U}) are explained in Appendix~\ref{Sec:SSandTsallis}.

\subsection{Entropy exchange with fluctuations described by gamma-distribution function}\label{Sec:Entropy1}
As is reviewed in Appendix~\ref{Sec:SSandTsallis}, Eq.~(\ref{rhob1}) can be written in terms of the auxiliary definition
\bea
\hvrr_b=\frac{1}{\mathcal{Z}}\exp_q\left(-\beta^\star\hH_b\right),
\label{varrho1}
\eea
where $\bs$ is a parameter related to the physical inverse temperature $\beta$ through:
\bea
\beta=\frac{\bs\,\text{Tr}\left[\hvrr^q_b(\bs)\right]}{1-(1-q) \bs \mathcal{U}\left(\bs\right) / \text{Tr}\left[\hvrr_b^q(\bs)\right]},
\label{Eq:RenormalizedBeta2}
\eea
with
\bea
\mathcal{U}=\text{Tr}\left[\hvrr^q_b\hH_b\right]=-\partial_{\bs}\ln_q\mathcal{Z},
\label{U2}
\eea
where
\bea
\mathcal{Z}=\text{Tr}\left[\exp_q\left(-\bs\hH_f\right)\right],
\label{Z2}
\eea
{ and the $q$-logarithm is defined as
\bea
\ln_q x=\frac{x^{1-q}-1}{1-q}.
\eea}

For a single-mode thermal state prepared according Eq.~(\ref{rhob1}) or Eq.~(\ref{varrho1}), it is straightforward to find that
\bea
p_n=\frac{\left[(q-1)\bs\omega\right]^{\frac{1}{q-1}}}{\zeta_\text{H}\left(\frac{1}{q-1},\frac{1}{(q-1)\bs\omega}\right)}\left[1-(1-q)n\bs\omega\right]^{\frac{1}{1-q}},
\eea
where $\zeta_\text{H}(s,x)$ is the Hurwitz-zeta function.

In order to define the entropy exchange, it is necessary to give an expression of the entropy for $\hrho_b(t)$. To do so, there are two options: the usual von-Neumann entropy
    \bea
    S_b(t)=-\text{Tr}\left[\hrho_b(t)\ln\hrho_b(t)\right],
    \label{vonNennumannentropy}
    \eea
 or the Tsallis-like entropy~\cite{Rajagopal2001}:
    \bea
    S_q^b(t)=-\text{Tr}\left[\hrho_b(t)\ln_q\hrho_b(t)\right].
    \eea

I will assume that the entropy is a continuous function on time. Therefore, if for $t=0$ the thermal state $\hrho_b$ is described with SS parameters such that the entropy follows a Tsallis-like form, then, for all $t$, the entropy prescription is the same. Also, it could be argued that information about the fluctuations (encoded in the $q$-parameter) needs to be present in the entropy definition at any instant of time. Hence, the entropy for the cavity mode will be taken as:
\bea
&&S_q^f(t)=-\left[p_0(1-\epsilon)+\mathcal{A}_0(t)\right]\ln_q\left[p_0(1-\epsilon)+\mathcal{A}_0(t)\right]\nn\\
&-&\left[\sum_{n=1}^\infty\left(\mathcal{A}_{n}(t)+\mathcal{C}_{n-1}(t)\right)\right]\ln_q\left[\sum_{n=1}^\infty\left(\mathcal{A}_{n}(t)+\mathcal{C}_{n-1}(t)\right)\right].\nn\\
\eea

The expressions reduce to Eq.~(\ref{vonNennumannentropy}) when $q\to1$.

{ In the same way, in order to have a unique definition of entropy for the whole system, the entropy of the atom is taken as with the Tsallis prescription:
\bea
&&S_a(t)=-\left[\sum_{n=0}^\infty\mathcal{A}_n(t)\right]\ln_q\left[\sum_{n=0}^\infty\mathcal{A}_n(t)\right]\nn\\
&-&\left[p_0(1-\epsilon)+\sum_{n=0}^\infty\mathcal{C}_{n}(t)\right]\ln_q\left[p_0(1-\epsilon)+\sum_{n=0}^\infty\mathcal{C}_{n}(t)\right],\nn\\
\label{Sat}
\eea}
and I define the partial entropy change:
\bea
\Delta S_j(t)\equiv S_j(t)-S_j(0),\quad j=a,b,
\label{PartialSexch}
\eea
and the sum of partial entropy changes:
\bea
\Delta\widetilde{S}(t)\equiv\Delta S_a(t)+\Delta S_b(t).
\label{Stilde}
\eea

It is clear that the functional form of the equations above depends on the parameter $\epsilon$, which carries information about the initial state of that subsystem. In order to explore different initial configurations, the density operator for the two-level system is parametrized in a Bloch form, namely:
\bea
\hrho_a(0)=\frac{1}{2}\left(\mathbb{1}+(2\epsilon-1)\hat{\sigma}_z\right),
\eea
so that, in the Bloch-sphere:
\bea
\epsilon=\frac{1+r\cos\theta}{2},
\label{epsilonpolar}
\eea
where $0\leq r\leq 1$ and $0\leq\theta\leq\pi$.

Finally, to give a time-independent understanding of the entropy, the partial average entropy exchange is computed as:%
\bea
\overline{\Delta S}_j=\frac{1}{T}\int_0^T \Delta S_j(t)dt,\quad j=a,b.
\label{SaverageDef}
\eea

\subsection{Entropy exchange with fluctuations described by a multi-level distribution function}\label{Sec:Entropy2}

For the density operator of Eq.~(\ref{rhoNlevel}), it is clear that
\bea
p_n=\frac{1}{Z_N}\sum_{k=1}^N e^{-n\beta_k\omega}.
\eea

{ And for this case, the entropy of the atom and cavity will be
\bea
&&S_a(t)=-\left[\sum_{n=0}^\infty\mathcal{A}_n(t)\right]\ln\left[\sum_{n=0}^\infty\mathcal{A}_n(t)\right]\nn\\
&-&\left[p_0(1-\epsilon)+\sum_{n=0}^\infty\mathcal{C}_{n}(t)\right]\ln\left[p_0(1-\epsilon)+\sum_{n=0}^\infty\mathcal{C}_{n}(t)\right],\nn\\
\eea
and
\bea
&&S_b(t)=-\left[p_0(1-\epsilon)+\mathcal{A}_0(t)\right]\ln\left[p_0(1-\epsilon)+\mathcal{A}_0(t)\right]\nn\\
&-&\left[\sum_{n=1}^\infty\left(\mathcal{A}_{n}(t)+\mathcal{C}_{n-1}(t)\right)\right]\ln\left[\sum_{n=1}^\infty\left(\mathcal{A}_{n}(t)+\mathcal{C}_{n-1}(t)\right)\right].\nn\\
\eea

To analyze the impact of the multi-level distribution function, the definitions given in Eqs.~(\ref{PartialSexch})-(\ref{SaverageDef}) remains valid.}

\begin{figure}[h!]
    \centering
    \includegraphics[scale=0.56]{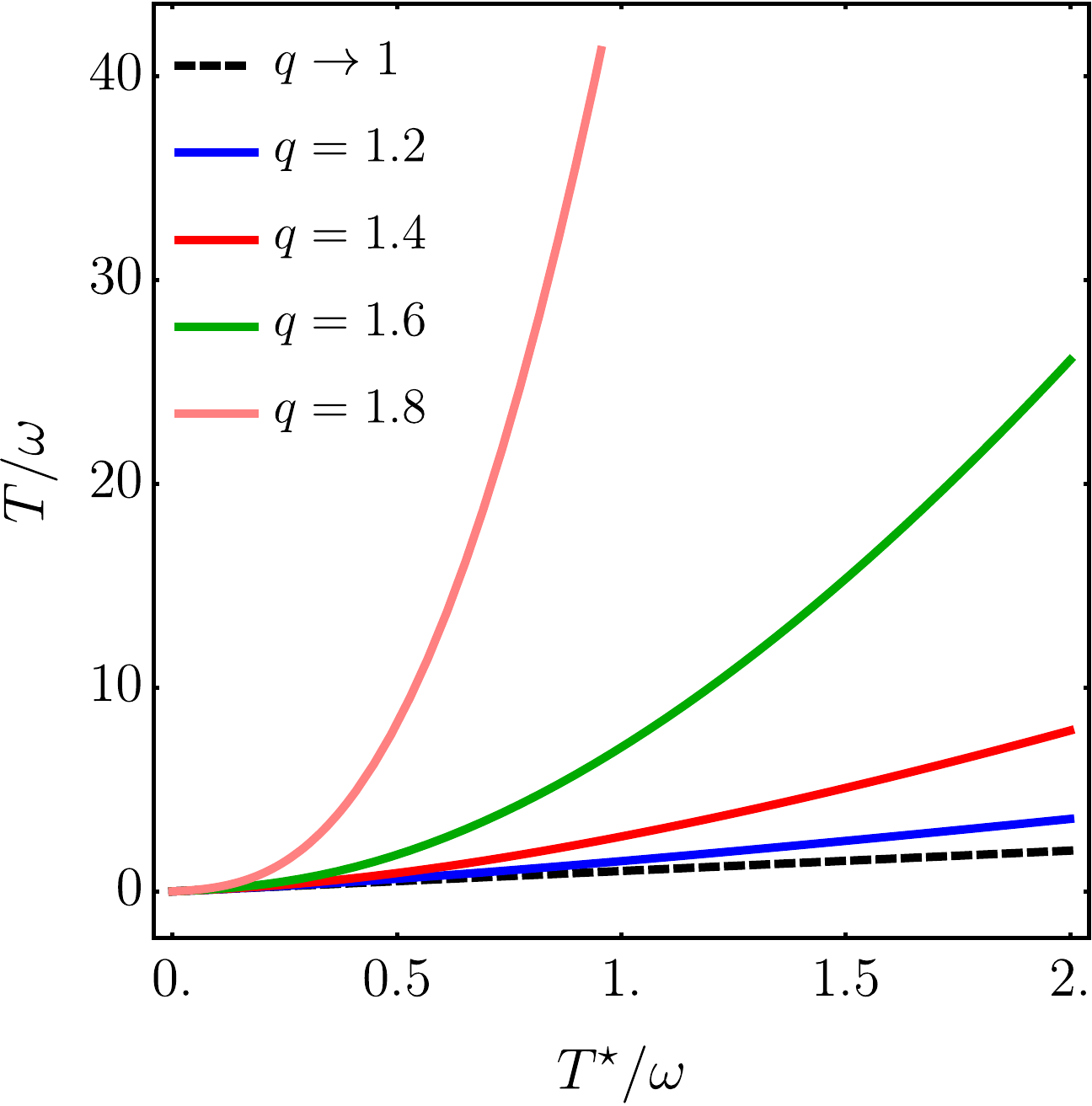}
    \caption{ Physical temperature $T=1/\beta$ as a function of the parameter $T^\star=1/\bs$ (both in units of $\omega$) computed from Eq.~(\ref{Eq:RenormalizedBeta2}). The Gibbs-Boltzmann limit, i.e., $q\to1$ (dashed line) is added for comparison.}
    \label{fig:physicalT}
\end{figure}

\section{Results and discussion}\label{Sec:Results}
\subsection{Fluctuations described by gamma-distribution function}\label{Sec:Results1}
\begin{figure}[h]
    \centering
    \includegraphics[scale=0.55]{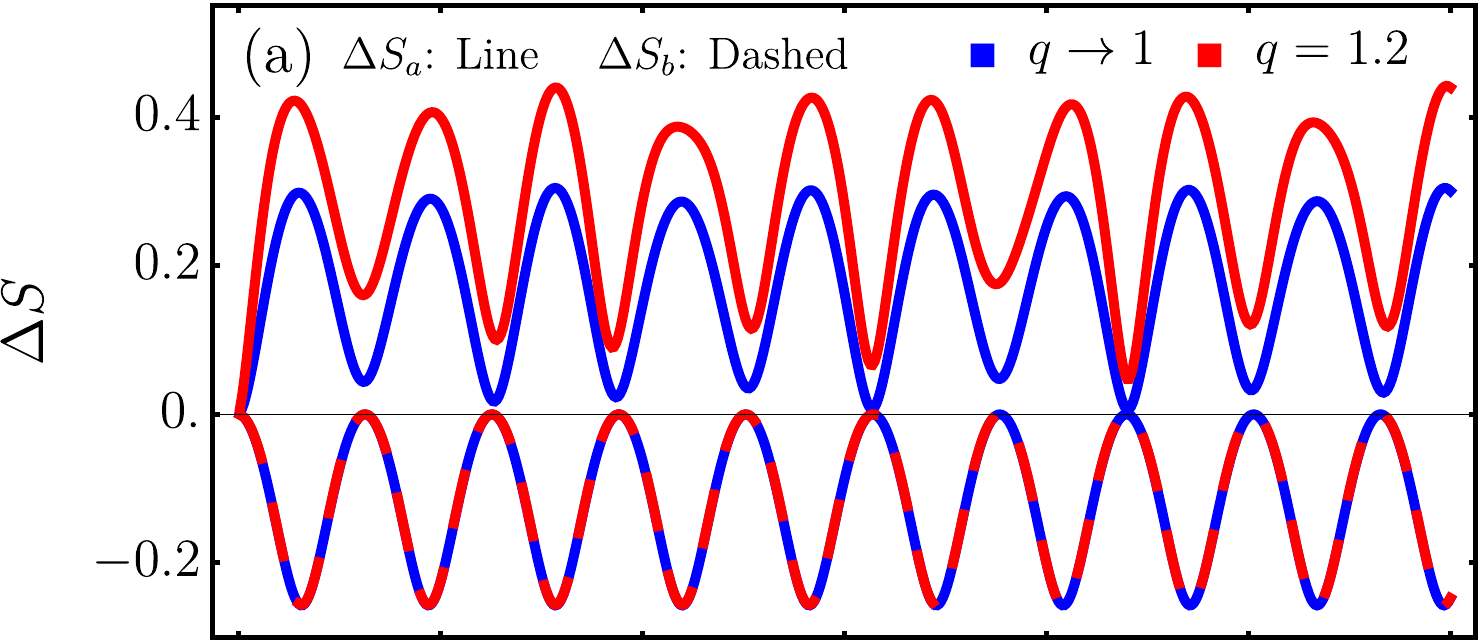}\\
    \includegraphics[scale=0.55]{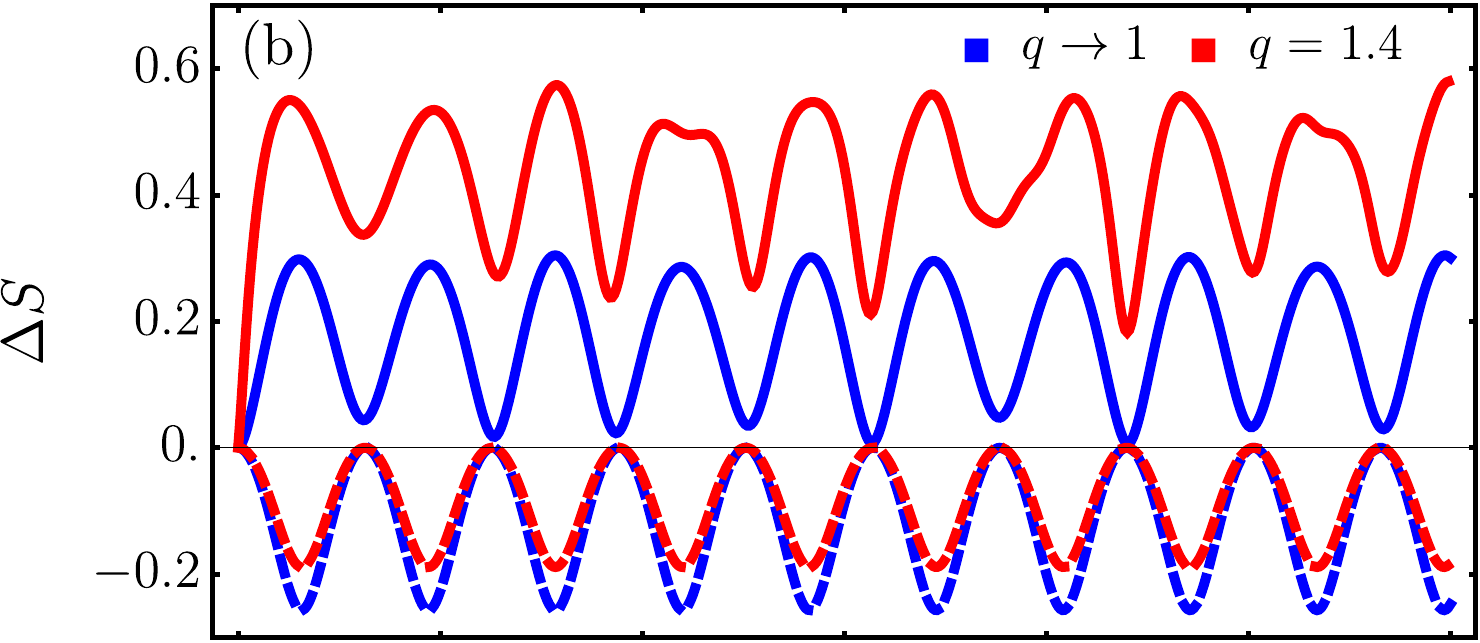}\\
    \includegraphics[scale=0.55]{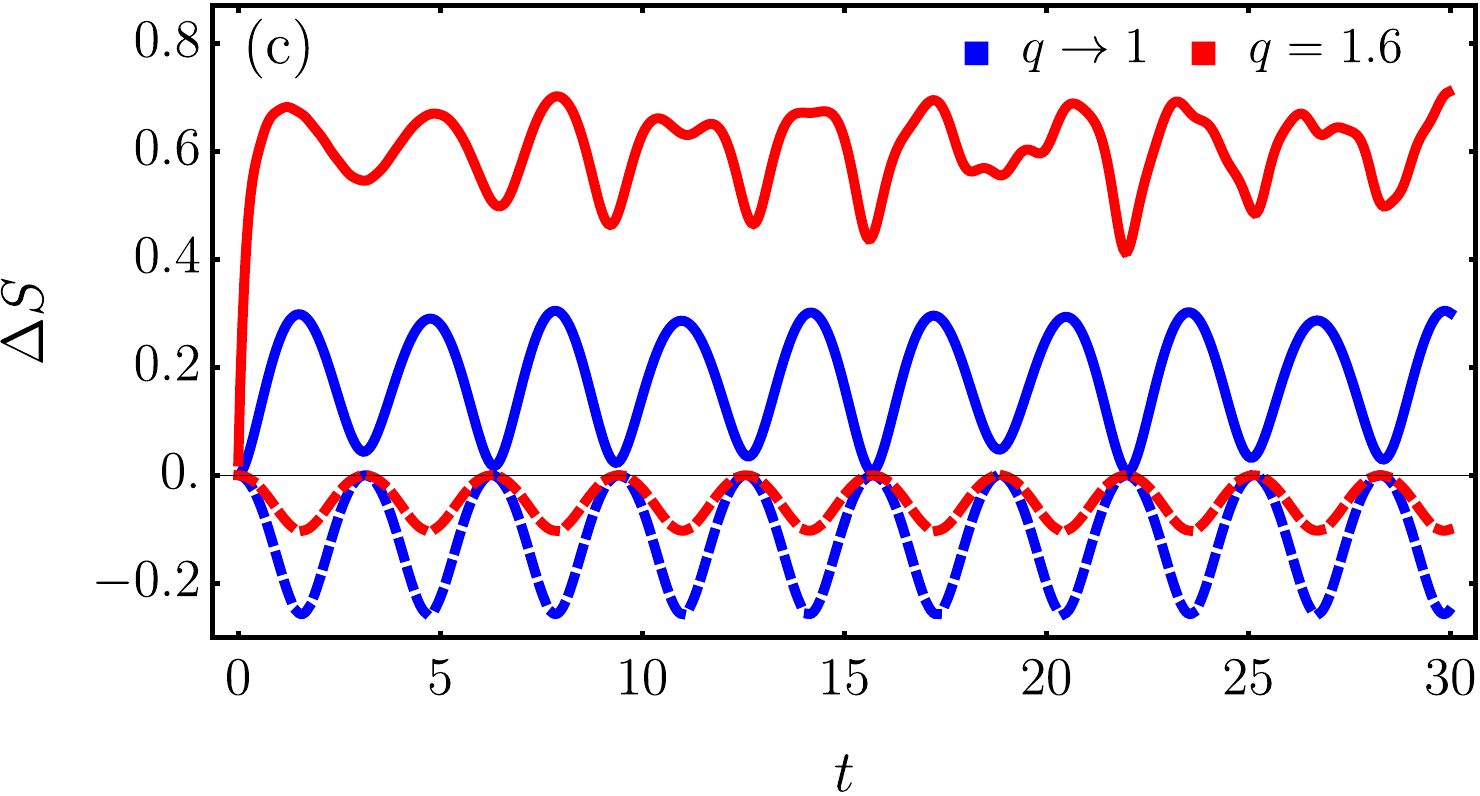}
    \caption{ Partial entropy exchange obtained from Eq.~(\ref{PartialSexch}) for the atom (line) and for the cavity mode (dashed), when the atom is prepared in its ground state ($\epsilon=0$). The different panels show the impact of the $q$-parameter (red) and the case $q=1$ (blue). The parameters are: $\Delta=0$, $\lambda=2$, and the physical temperature is fixed to the value $\beta\omega\approx2.39$ for which $\bar{n}=0.1$ when $q=1$.}
    \label{fig:DeltaSdifq}
\end{figure}

In order to use model parameters connected with the well-known phenomenology, the values of $q$ and $\beta\omega$ are restricted to lie in a certain interval. The latter is because, for all those values, the physical temperature of Eq.~(\ref{Eq:RenormalizedBeta2}) cannot be defined, which is connected with the restriction on the $q$-exponential, where $1+(1-q)x\geq0$. This effect is present in kappa-distributions where the non-equilibrium dynamics of correlated systems imply that temperature is bounded below, i.e., the information about the thermal state is not simultaneously shared for all the subsystem parts~\cite{livadiotis2015introduction,livadiotis2009beyond,livadiotis2013evidence}. Moreover, recently, that bound to a physical temperature definition was found in a spin-$1/2$ { $XX$ dimer model}~\cite{castano2021super}. In the following, the non-extensive parameter will take the values $1< q< 2$, which is a range where the parameters of the Hamiltonian of Eq.~(\ref{Hamiltonian}) allow a less restricted physical temperature definition. However, the formalism presented in this work can be easily extrapolated to other values of $q$ with their corresponding temperature interval. Figure~\ref{fig:physicalT} shows the deviations of the physical $T$ from the Gibbs-Boltzmann limit ($q\to1$), when the parametrization with $T^\star=1/\bs$ is performed. As can be noticed, the physical temperature related to the cavity mode for $q\neq1$ is higher than the one without thermal fluctuations.

\begin{figure}[h]
    \centering
    \includegraphics[scale=0.55]{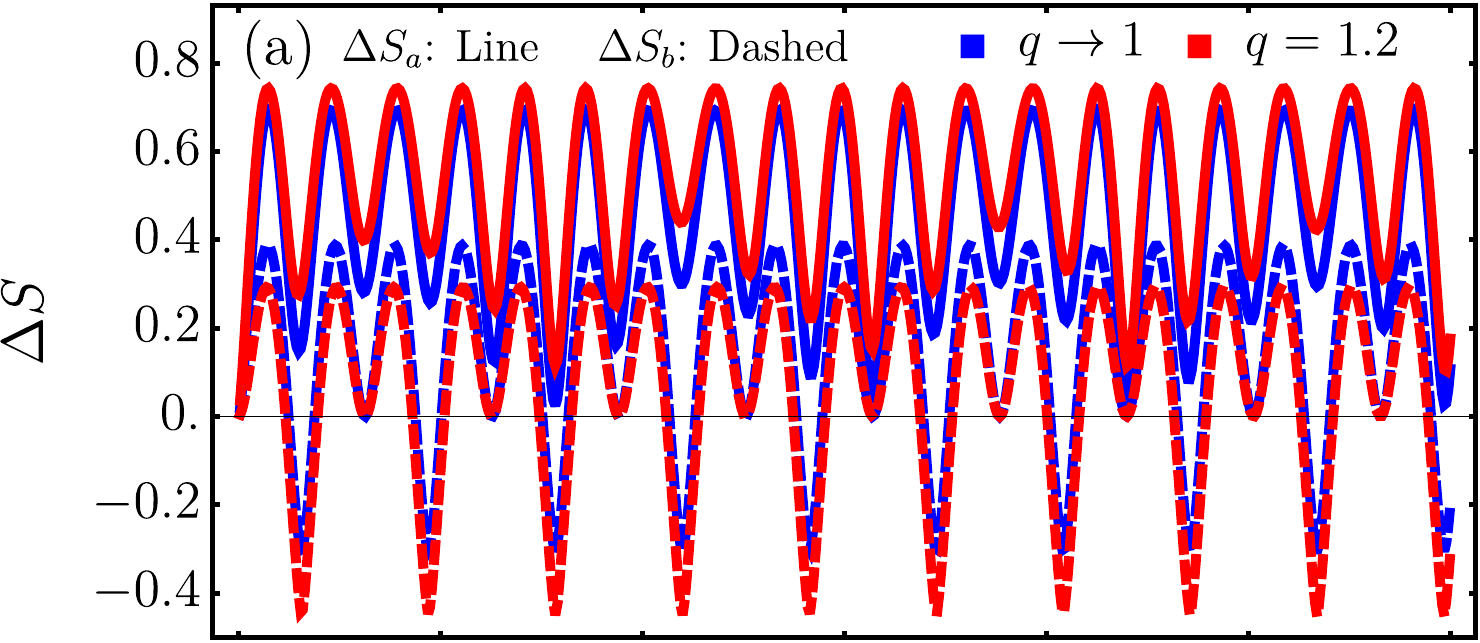}\\
    \includegraphics[scale=0.55]{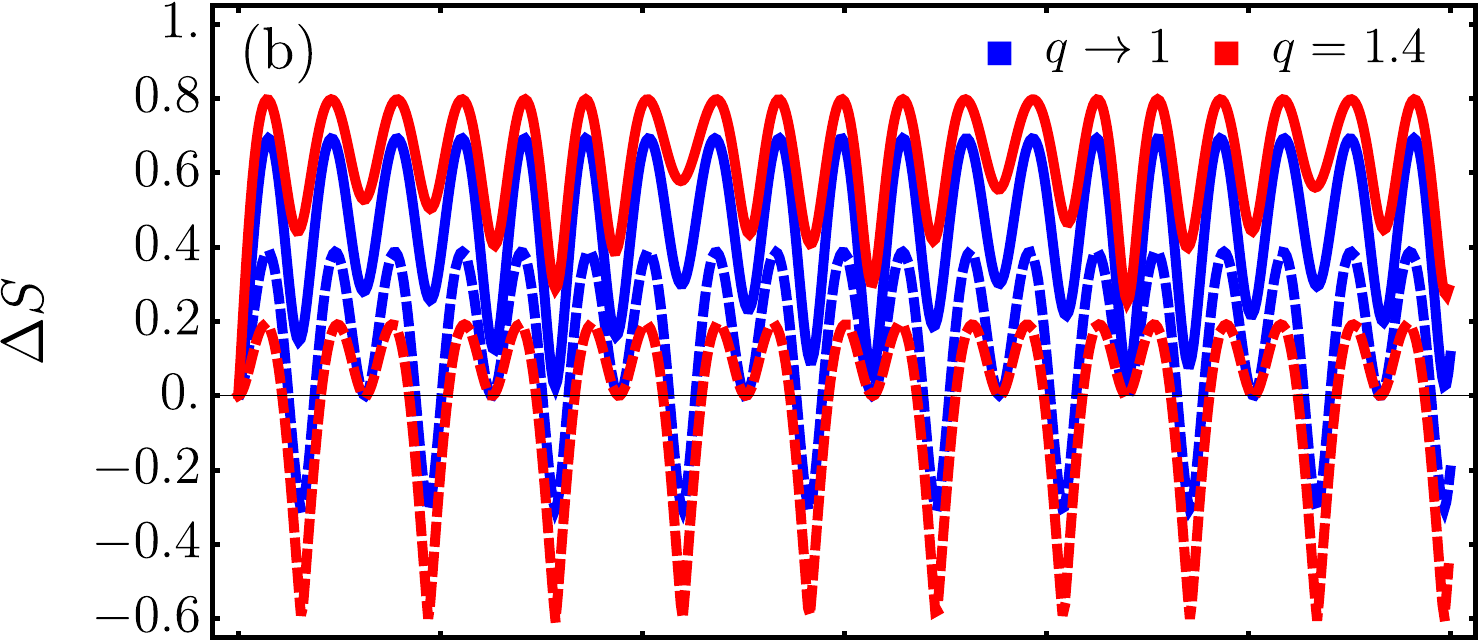}\\
    \includegraphics[scale=0.55]{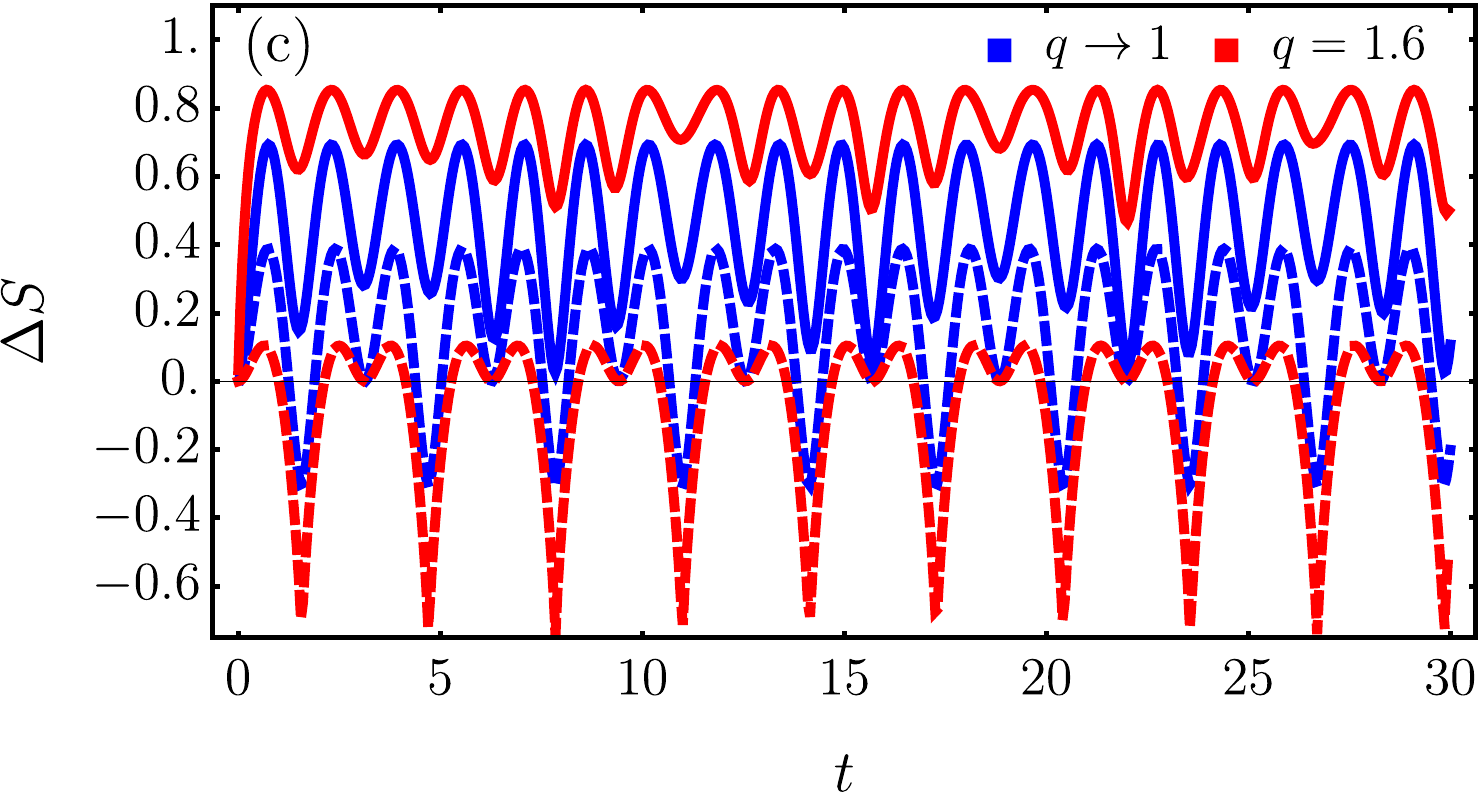}
    \caption{The same of Fig.~\ref{fig:DeltaSdifq} but when the atom is prepared in its excited state ($\epsilon=1$).}
    \label{fig:DeltaSdifqep1}
\end{figure}

Figure~\ref{fig:DeltaSdifq} shows the partial entropy exchange as a function of time for the atom (continuous lines) and the cavity mode (dashed lines), when the parameter $q$ has the values $q=1, 1.2, 1.4$, and $1.6$. In that figure, the parameters of the model are: $\Delta=0$, $\lambda=2$, and the initial state of the atom is set to $\hrho_a(a)=\ket{g}\bra{g}$, i.e., $\epsilon=0$. For all the panels the inverse temperature is set to $\beta\omega\approx2.39$, which is the inverse temperature of a weakly excited thermal field ($\bar{n}=0.1$) when $q\to1$. As the figure indicates, both the atom and the field suffer substantial modifications in their entropy exchange when $q$ rises. Moreover, besides the amplitude of each one, the functional behavior of $\Delta S_a(t)$ is highly modified, whereas $\Delta S_b(t)$ remains close to a sinusoidal form.

It is worth mentioning that even if the inverse temperature was fixed for each situation described above ($\beta\omega\approx2.39$), the average number of photons in the cavity is not the same. In fact, as presented in Appendix~\ref{Sec:Naverage}, the $q$-average number of photons, $\bar{n}_q$, is:
\bea
\bar{n}_q
&=&\frac{1}{\zeta_\text{H}\left(\frac{q}{q-1},\frac{1}{(q-1)\bs\omega}\right)}\Bigg[\Phi\left(1,\frac{1}{q-1},\frac{1}{(q-1)\bs\omega}\right)\nn\\
&-&\frac{1}{(q-1)\bs\omega}\Phi\left(1,\frac{q}{q-1},\frac{1}{(q-1)\bs\omega}\right)\Bigg],
\eea
where $\Phi(z,s,r)$ is the so-called Hurwitz-Lerch transcendent function~\cite{zwillinger2007table}, defined as
\bea
\Phi(z,s,r)=\sum_{n=0}^\infty\frac{z^n}{(n+r)^s}.
\eea

Therefore, for $\beta\omega\approx2.39$,
\bea
\bar{n}_{1.2}&\approx&0.102773,\nn\\ \bar{n}_{1.4}&\approx&0.100935,\nn\\
\bar{n}_{1.6}&\approx&0.094662,
\eea
so that the field is still in a weakly excited thermal state.

\begin{figure}[h]
    \centering
    \includegraphics[scale=0.55]{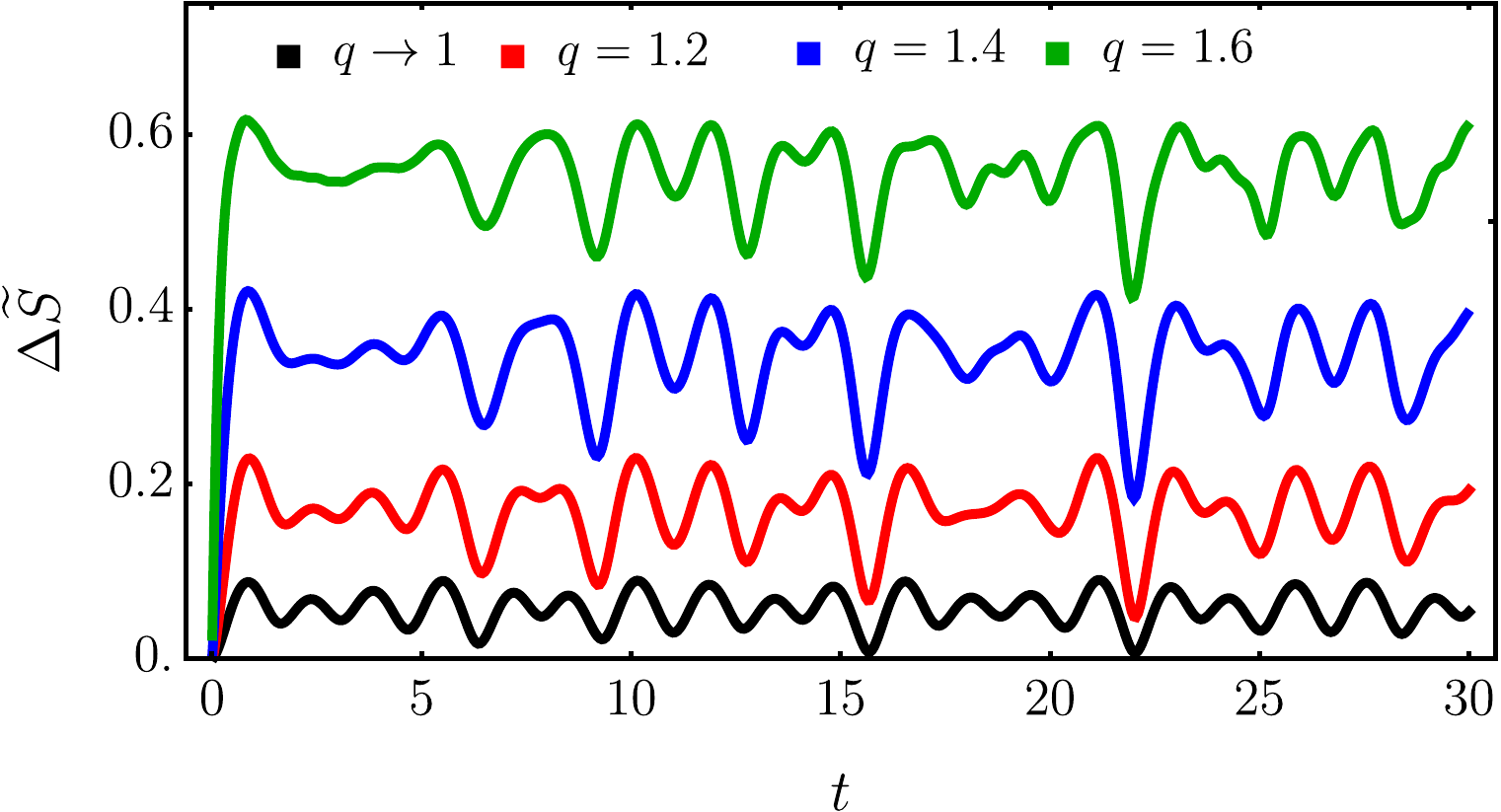}
    \caption{Sum of the atomic and field partial entropy changes given in Eq.~(\ref{Stilde}) for $q\to 1$ (black), $q=1.2$ (red), $q=1.4$ (blue) and $q=1.6$ (green). The parameters are the same of Fig.~\ref{fig:DeltaSdifq}.}
    \label{fig:DeltaStilde}
\end{figure}
\begin{figure}[h]
    \centering
    \includegraphics[scale=0.53]{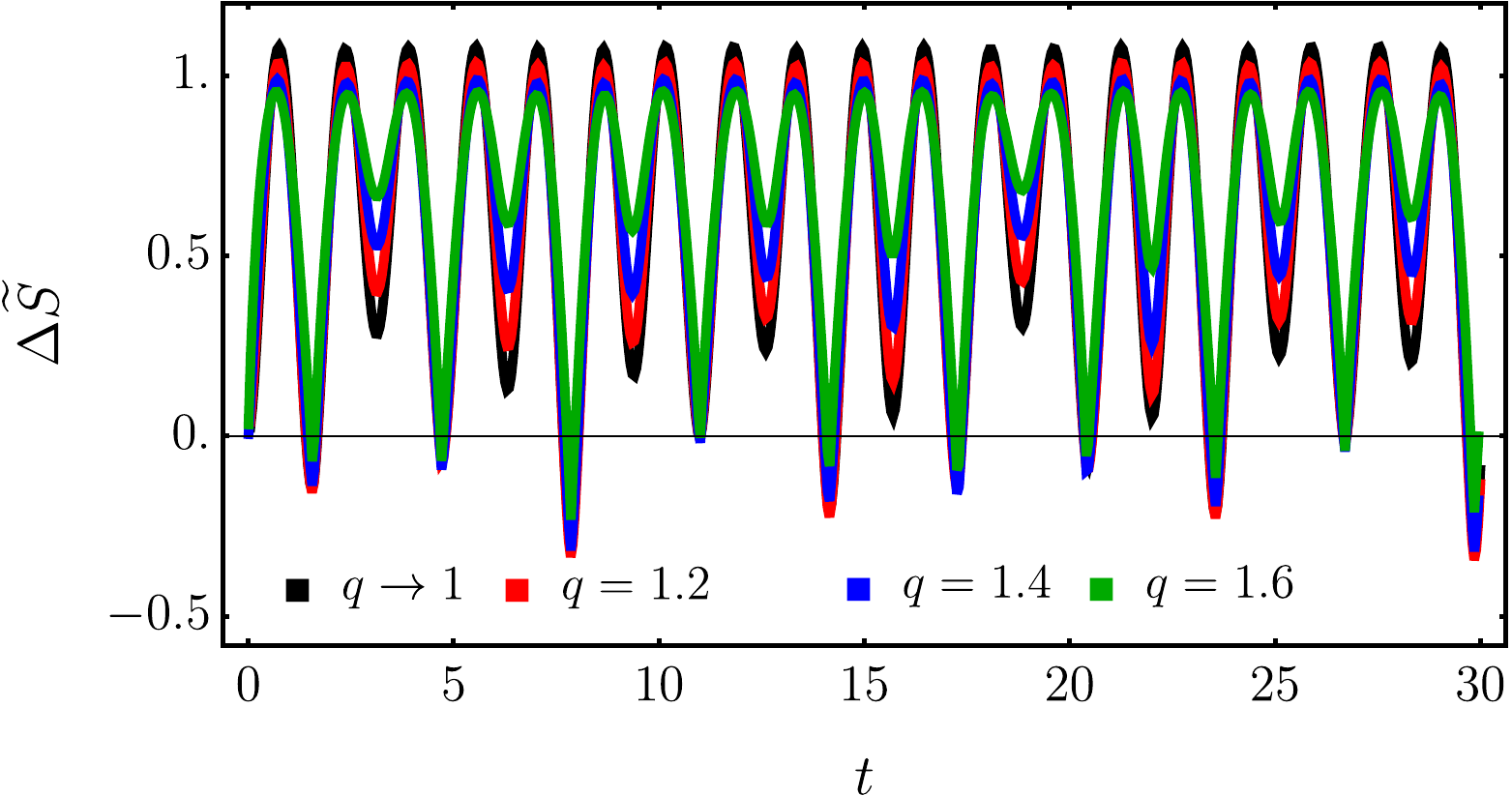}
    \caption{The same of Fig.~\ref{fig:DeltaStilde} but when the atom is prepared in its excited state ($\epsilon=1$).}
    \label{fig:DeltaStildeep1}
\end{figure}

\begin{figure*}[t]
    \centering
    \includegraphics[scale=0.43]{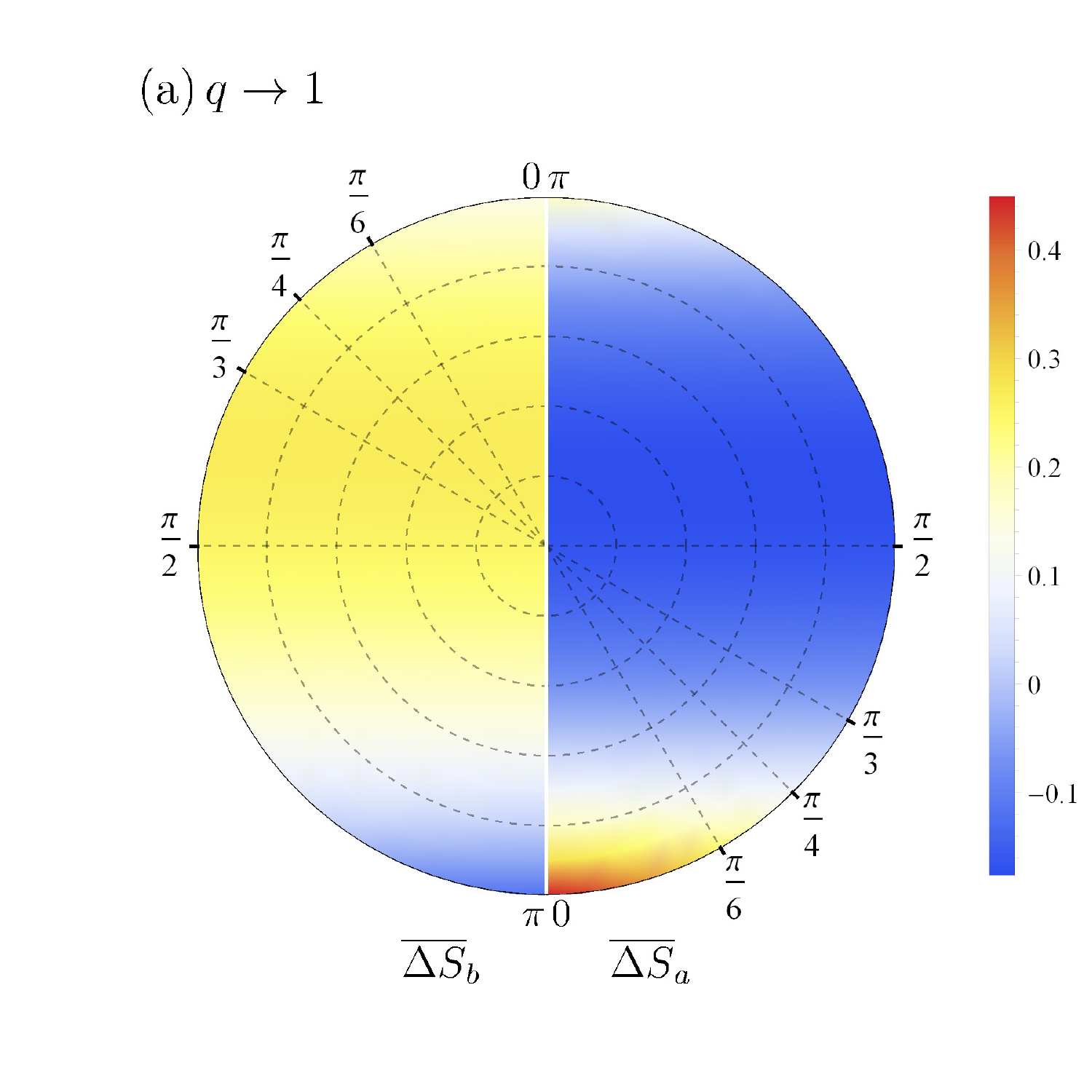}\hspace{0.5cm}\includegraphics[scale=0.43]{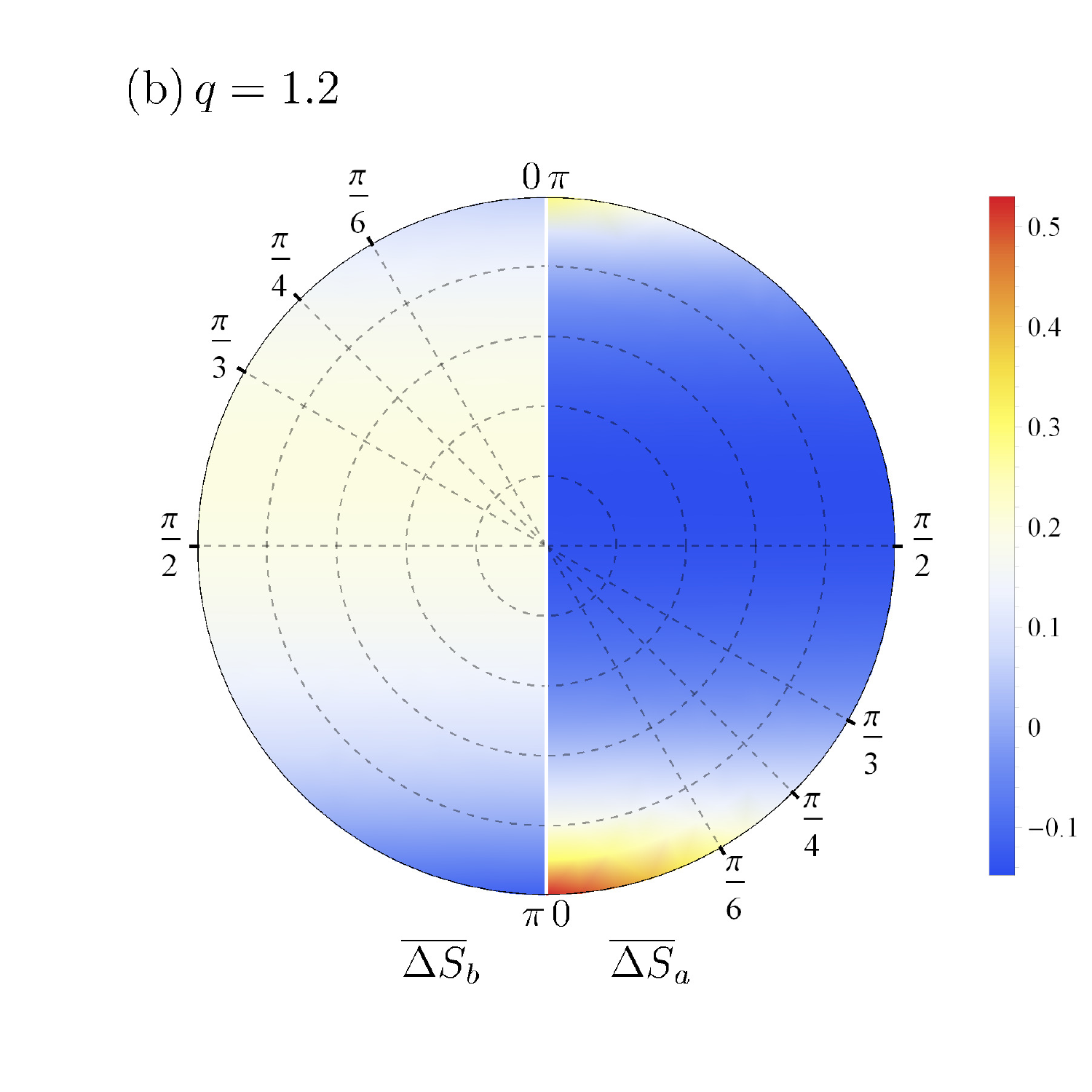}\\
    \vspace{0.3cm}
        \includegraphics[scale=0.43]{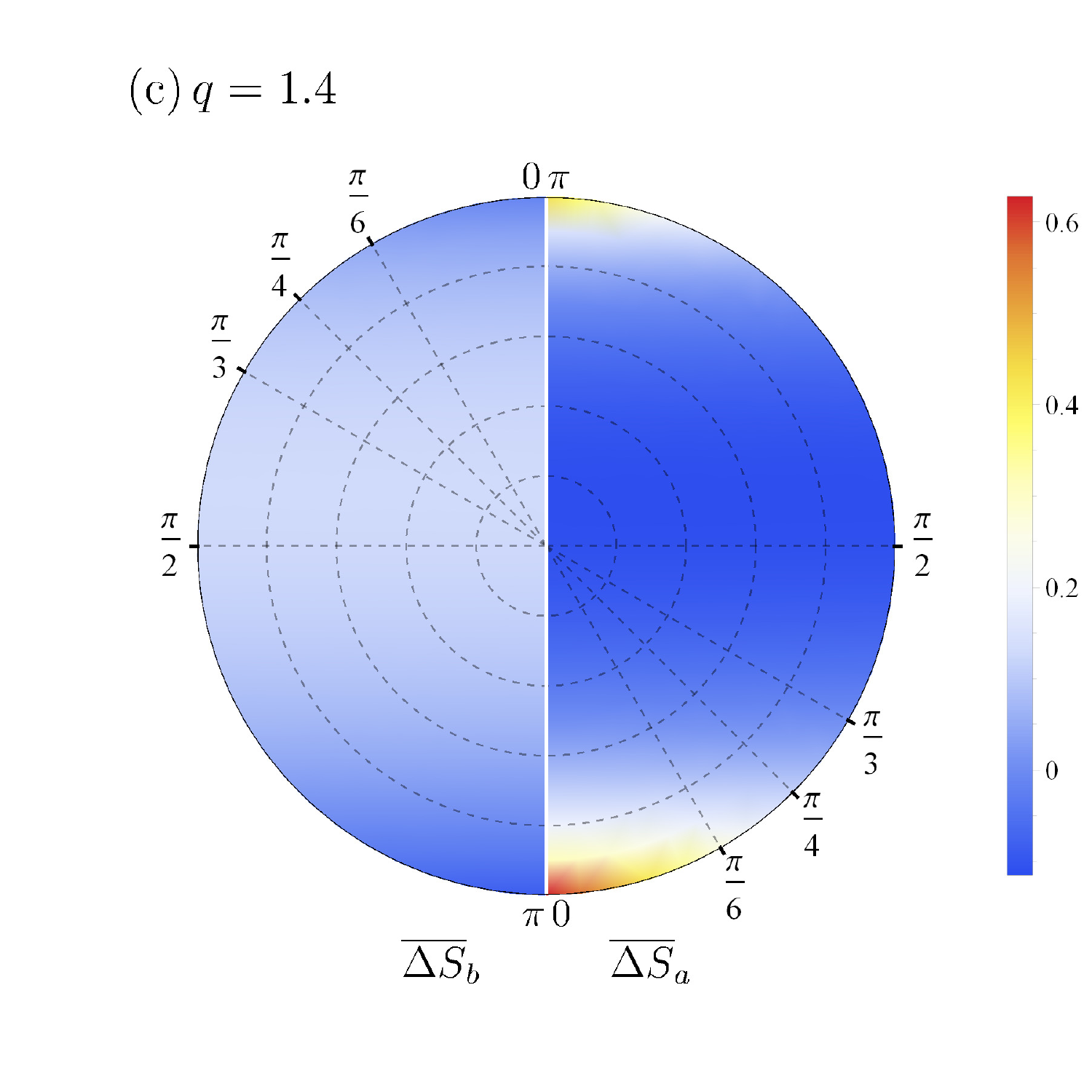}\hspace{0.5cm}\includegraphics[scale=0.43]{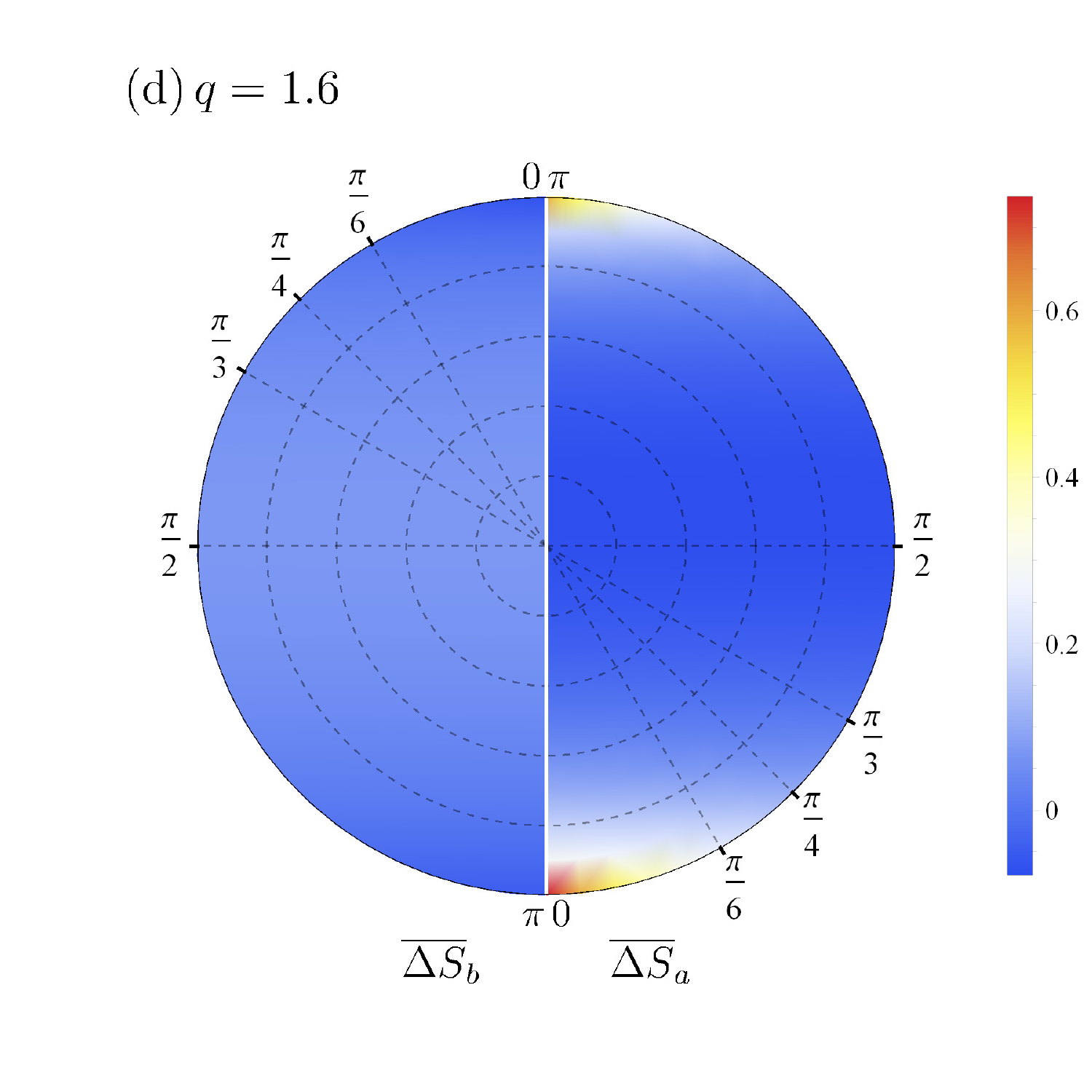}
    \caption{Partial average entropy exchange of Eq.~(\ref{SaverageDef}) as a function of $\epsilon$, represented in the polar Bloch sphere parametrized with $(r,\theta)$ by following Eq.~(\ref{epsilonpolar}). In each panel, the left-side represent $\overline{\Delta S}_b$, whereas the right-side is $\overline{\Delta S}_a$. The dashed circumferences represent the regions for which $r=0.2,0.4,0.6$ and $0.8$, whereas the radial dashed lines gives the notable angles $\theta= 0,\pi/6,\pi/4,\pi/3,\pi/2$ and $\pi$.}
    \label{fig:DensityS1}
\end{figure*}

Figure~\ref{fig:DeltaSdifqep1} shows the time-variation of the exchange entropy, when $\hrho_a(0)=\ket{a}\bra{a}$ or $\epsilon=1$. In this case, $\Delta S_a(t)$ and $\Delta S_b(t)$ exhibit changes in their amplitude but their function form remains unaltered. { Compared with $q\to1$, both $\Delta S_a(t)$ and $\Delta S_b(t)$ considerably move away from the Gibbs-Boltzmann limit.} In Figs.~\ref{fig:DeltaStilde} and~\ref{fig:DeltaStildeep1} the sum of partial entropies is plotted for $\epsilon=1$ and $\epsilon=1$, respectively. As found in Figs.~\ref{fig:DeltaSdifq} and~\ref{fig:DeltaSdifqep1}, the more dramatic changes in $\Delta S_j$ come from when the atom is prepared in its ground state. Note that for $\epsilon=0$, $\Delta\widetilde{S}(t)$ enhances when $q$ grows, indicating that the entropy exchange increases if the cavity mode is far from the thermal equilibrium. On the contrary, for $\epsilon=1$, the total entropy exchange remains close to the thermal equilibrium situation, and its value decreases when $q$ grows.

For the sake of completeness, Fig.~\ref{fig:DensityS1} shows the time-averaged entropy exchange of Eq.~(\ref{SaverageDef}) for different values of $\epsilon$ represented in the Bloch-sphere parametrization given in Eq.~(\ref{epsilonpolar}). As Fig.~\ref{fig:DensityS1}-(a) indicates, in the Gibbs-Boltzmann limit, $\Delta S_a$ and $\Delta S_b$ (in average) are positive and negative, respectively, near to $r=1$ and $\theta=\pi$ ($\epsilon=0$). That situation remains for $q\neq 1$. However, although for different $\epsilon$ the time-average for $\Delta S_a$ remains almost unaltered, the value of $\overline{\Delta S}_b$ decreases for regions where $0\leq r\leq1$ and around $0<\theta<3\pi/4$. In other words, for the associated $\epsilon$ in those regions, the average entropy exchange for $t>0$ is less than the entropy of the initial state of the cavity mode. { Moreover, the region in the equilibrium limit $q=1$ for which $\overline{\Delta S}_a$ reaches its maximum values ($r\sim 1, 0<\theta\lesssim\pi/6$) is diminished so that for $q=1.6$, $r\sim 1$ and $0<\theta\lesssim\pi/12$. The latter indicates that the initial preparation of the atom is important in the presence of thermal fluctuations.}

{ In the light of the results, it is necessary to comment that the $q$-parameter is not just a scaling factor without a physical interpretation. In fact, as is pointed out in Ref.~\cite{beck2003superstatistics}, in the SS context, $q$ is related to the standard deviation $\sigma$ of the probability density function as:
\bea
(q-1)\beta^2=\sigma^2\text{ or } q=\frac{\langle\tilde{\beta}^2\rangle}{\langle\tilde{\beta}\rangle^2},
\eea
where $\beta$ is the average inverse temperature and $\tilde{\beta}$ is the fluctuating inverse averaged temperature. Then, $q$ provides information about how significant the thermal fluctuations are, and how the system differs from the Gibbs-Boltzmann thermodynamics.}

\subsection{Fluctuations described by a two-level distribution function}\label{Sec:Results2}
\begin{figure}[h!]
    \centering
    \includegraphics[scale=0.53]{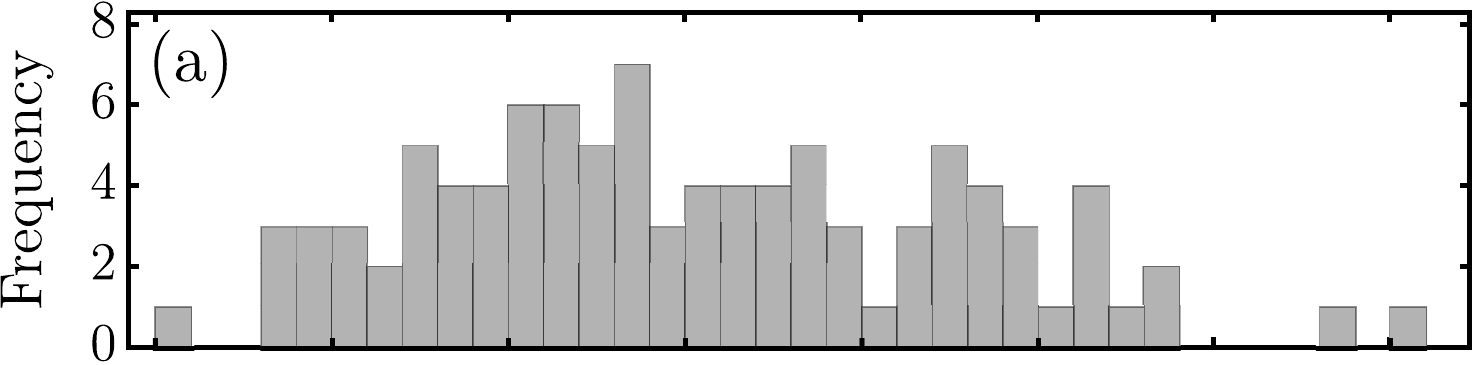}\\
    \vspace{0.3cm}
    \!\includegraphics[scale=0.53]{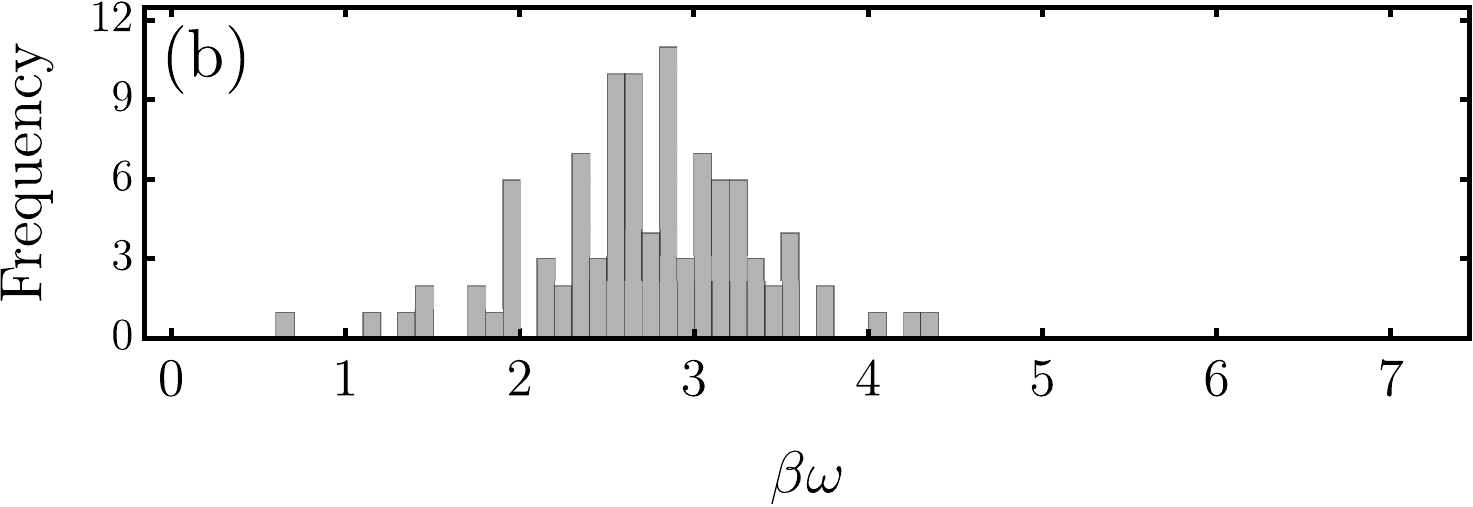}
    \caption{Random valued of $\beta\omega$ around $\beta\omega=3$, generated by following (a) normal distribution, and (b) Weibull distribution.}
    \label{fig:Distbeta}
\end{figure}

In order to study a cavity mode where the fluctuations are spread over a finite number of temperatures, I set $N=100$ and the values of $\beta\omega$ were generated randomly, by following different distribution shapes around $\beta\omega=3$. Figure~\ref{fig:Distbeta} shows the inverse temperature distribution used, related to normal and Weibull distributions.

\begin{figure}[h!]
    \centering
    \includegraphics[scale=0.53]{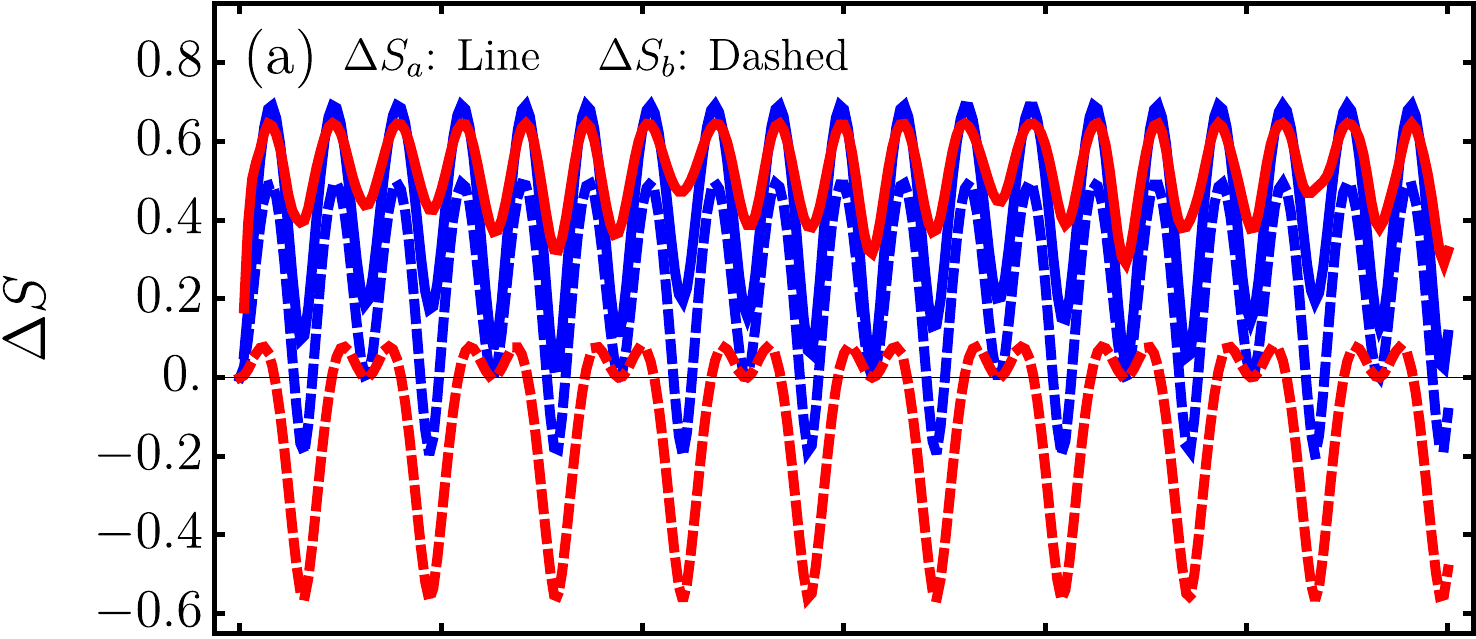}\\
    \includegraphics[scale=0.53]{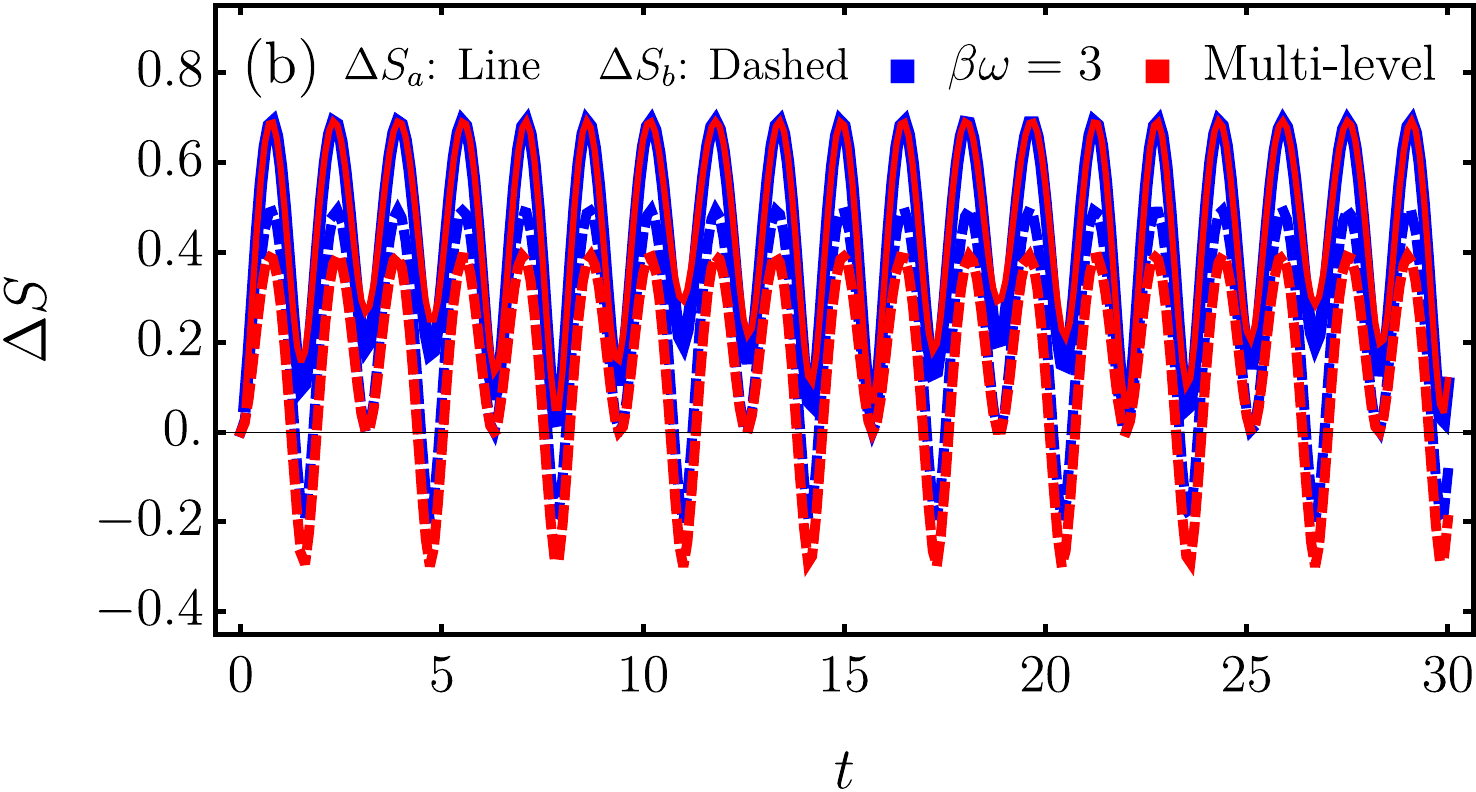}
    \caption{Partial entropy exchange obtained from Eq.~(\ref{PartialSexch}) for the atom (line) and for the cavity mode (dashed), when the atom is prepared in its excited state ($\epsilon=1$). Color blue indicates the Gibbs-Boltzmann limit with $\beta\omega=3$, whereas red is for the multi-level distribution: (a) Normal and (b) Weibull. The parameters are $\Delta=0$ and $\lambda=2$.}
    \label{fig:MultivelelDeltaS}
\end{figure}

In Fig.~\ref{fig:MultivelelDeltaS}, the partial entropy exchange is shown, for the two distributions of aleatory inverse temperatures. The parameters of the model are $\Delta=0$, $\lambda=2$ and $\epsilon=1$ (the atom is initially in its excited state). In Fig.~\ref{fig:MultivelelDeltaS}-(a), the effect of the multi-level distribution function is to modify the amplitude of $\Delta S_a(t)$, but maintaining its functional shape (continuous lines). The multi-level approximation (red) gives an enhanced partial entropy exchange compared to the Gibbs-Boltzmann limit, where $\beta\omega=3$. On the other hand, the cavity mode partial entropy exchange (dotted lines) has a inverse behavior: the selected distribution makes that $\Delta S_b(t)$ becomes more negative (red) compared with the single temperature case (blue). Also, local minima are strongly modified.

In the Weibull distribution, as Fig.~\ref{fig:MultivelelDeltaS}-(b) shows, the partial entropy exchange for the atom and field is slightly modified. Then, the way the inverse temperature distributes is determinant to get appreciable deviations.

\section{Summary and Conclusions}\label{Sec:Concl}
This work studied the impact of the thermal fluctuations on the entropy exchange of an atom and a single cavity mode within the Jaynes-Cummings model. The SS framework was used with the gamma and the multi-level distribution functions to model the out-of-equilibrium situation. For the first one, restrictions over the free parameters parallel the Tsallis entropy theory. In this case, and to ensure a physical definition of temperature, besides the Legendre structure preservation in thermal observables, constraints to energy are imposed so that the thermal state of the cavity mode can be computed analytically by introducing a parametrization of the temperature. { The Legendre structure preservation implies that the thermodynamic functions need to be computed with the $q$-logarithm prescription, so that $q$ can be identified with a parameter that carries information about the thermal fluctuations in connection with the particular choice of the temperature distribution. The values of the free parameters of the Hamiltonian give additional restrictions over the $q$-parameter and the definition of the physical temperature, which  is found greater than the usual limit $q\to1$ when $1<q<2$. Moreover, all the results were performed with the same temperature, implying that the mean occupation number for different $q$ varies.}

 As discussed in Sec.~\ref{Sec:Results1}, the time-dependent partial entropy changes are more pronounced when the $q$-parameter moves away from the Gibbs-Boltzmann limit. In addition to the non-additive effect of the cavity mode (controlled by the parameter $\epsilon$), the initial state of the atom is crucial to obtain a positive or negative average partial $q$-dependent entropy for each subsystem. Then, the exchange of information for the atom and field is determined by fine tuning the values of $\epsilon$ and $q$.

On the other hand, for the case of a multi-level distribution function, it was found that the way how several temperatures are distributed is important to obtain deviations from the thermal equilibrium configuration. In particular, if $N$-values of inverse temperature were randomly obtained by following a normal distribution, appreciable differences appear when the results are compared with the single-inverse temperature case. Nonetheless if the random temperatures are assigned with a Weibull distribution, the time-dependent entropy exchange is slightly modified. 

{ It is essential to recall that the framework presented in this article is the direct application of an {\it ansatz}, and therefore, it does not constitute a microscopic or fundamental description of the non-equilibrium situation. The latter is connected with the fact that all the analysis is related to a single parameter ($q$ or $\beta_k$) that carries information about the thermal fluctuations. However, the $q$-parameter may be reached from several experimental setups as long as the thermal fluctuations are prepared by following Eq.~(\ref{chisquared}) with the corresponding constrains for $b$ and $c$. The same applies to the case of the multi-level approximation. Therefore, even when the current discussion is about the impact of the change of the parameters, it could model several physical scenarios.}

In conclusion, depending on the model adopted for the treatment of the temperature fluctuations, { (i.e. the way of $\beta_k$'s are distributed or the value of $q$), and the initial state of the atom and cavity (the values of $\epsilon$), observables of interest like the total, and  partial entropy or the entanglement change in a non-trivial way}. 

\section*{Acknowledgements}
The author acknowledges support from Consejo Nacional de Ciencia y Tecnolog\'ia CONACyT (M\'exico) under grant number A1-S-7655. Also, the author thanks  Emiliano Adrián Rodríguez Reyes, and Carolina Tavares for a thorough reading of the manuscript and the language correction. Also, I thank Dr. Edgar Guzmán for his valuable comments. 
\appendix
\section{Super-Statistics and Tsallis thermodynamics}\label{Sec:SSandTsallis}
The connection of SS and Tsallis thermodynamics comes from the entropy generalization for  non-additive systems~\cite{tsallis1988possible}, namely:
\bea
S=\frac{1}{q-1}\left(1-\text{Tr}\left[\hrho^q\right]\right)\forall\;q \in\mathbb{R},
\label{TsallisEntropy}
\eea
where $q$ is the non-additive index and $\hrho$ is the density operator of the system. In order to mimic the main results of the Boltzmann statistics, it is argued that the Tsallis framework has to full fill the so-called Legendre structure of thermodynamics~\cite{tsallis1998role}, which can be demanded by fundamental arguments related to an increasing entropy and a positive-definite specific heat~\cite{plastino1997universality,scarfone2016consistency}. In order to implement that idea, the internal energy $U$ is constrained by the following average-prescription:
\bea
U=\frac{\text{Tr}\left[\hrho^q\hat{H}\right]}{\text{Tr}\left[\hrho^q\right]},
\label{Uapp}
\eea
where $\hat{H}$ is the Hamiltonian. Then, by maximizing the functional associated to $S$ and $U$, the density operator is given by:
\bea
\hrho=\frac{1}{Z}\exp_q\left(-\beta\frac{\hat{H}-U}{\text{Tr}\left[\hrho^q\right]}\right),
\label{rhoapp}
\eea
where $\beta$ is the inverse physical temperature and the partition function $Z$ is given by
\bea
Z=\text{Tr}\left[\exp_q\left(-\beta\frac{\hat{H}-U}{\text{Tr}\left[\hrho^q\right]}\right)\right].
\label{Zapp}
\eea

The Eqs.~(\ref{Uapp})-(\ref{Zapp}) are implicit for $\hrho$ and their solution is not trivial. Nevertheless, there is an auxiliary form to avoid that problem. If an auxiliary density matrix $\hvrr$ is defined as
\bea
\hvrr=\frac{1}{\mathcal{Z}}\exp_q\left(-\beta^\star\hat{H}\right),
\label{rho2app}
\eea
where
\bea
\mathcal{Z}=\text{Tr}\left[\exp_q\left(-\beta^\star\hat{H}\right)\right].
\label{Z2app}
\eea
and $\bs$ is a quasi-temperature parameter given by
\bea
\beta=\frac{\beta^\star\,\text{Tr}\left[\hvrr^q(\beta^\star)\right]}{1-(1-q) \beta^\star \mathcal{U}\left(\beta^\star\right) / \text{Tr}\left[\hvrr^q(\beta^\star)\right]},
\label{Eq:RenormalizedBeta2pp}
\eea
the following relation is found:
\bea
\hrho(\beta)&=&\hvrr(\beta^\star),
\label{Z1andZ}
\eea
where $\mathcal{U}$ is defined in Eq.~(\ref{U2}). Therefore, the problem for $\hrho$ is solved by working with $\hvrr$ and with Eqs.~(\ref{Z1andZ}), together with the physical temperature parametrization of Eq.~(\ref{Eq:RenormalizedBeta2pp}).
\section{The non-additive average photon number}\label{Sec:Naverage}
For the non-extensive case, the average photon number is given by
\bea
\bar{n}_q=\frac{\text{Tr}_R\left[\hvrr^q_R(\bs)\hat{n}\right]}{\text{Tr}_R\left[\hvrr^q_R(\bs)\right]}=\frac{1}{\text{Tr}_R\left[\hvrr^q_R(\bs)\right]}\sum_{n=0}^\infty n\left(1+\alpha n\right)^{\nu},\nn\\
\eea
where $\alpha\equiv-(1-q)\bs\omega$ and $\nu\equiv q/(1-q)$. 

In order to compute the sum above, let me introduce the regulator $\eta$ in the following way:
\bea
\mathcal{S}(\alpha,\nu)&=&\sum_{n=0}^\infty n\left(1+\alpha n\right)^{\nu}\nn\\
&=&-\lim_{\eta\to0}\frac{\partial}{\partial\eta}\sum_{n=0}^\infty \left(1+\alpha n\right)^{\nu}e^{-\eta n},
\eea
but
\bea
\sum_{n=0}^\infty \left(1+\alpha n\right)^{\nu}e^{-\eta n}=\alpha^\nu\Phi\left(e^{-\eta},-\nu,\frac{1}{\alpha}\right),
\eea
where $\Phi(z,s,r)$ is the so-called Hurwitz-Lerch transcendent function, defined as
\bea
\Phi(z,s,r)=\sum_{n=0}^\infty\frac{z^n}{(n+r)^s}.
\eea

Then,
\bea
\mathcal{S}(\alpha,\nu)&=&-\alpha^\nu\lim_{\eta\to0}\frac{\partial}{\partial\eta}\Phi\left(e^{-\eta},-\nu,\frac{1}{\alpha}\right)\nn\\
&=&\alpha^\nu\left[\Phi\left(1,-1-\nu,\frac{1}{\alpha}\right)-\frac{1}{\alpha}\Phi\left(1,-\nu,\frac{1}{\alpha}\right)\right].\nn\\
\eea

Therefore,
\bea
\bar{n}_q&=&\frac{\left[(q-1)\bs\omega\right]^{\frac{q}{1-q}}}{\text{Tr}_R\left[\hvrr^q_R(\bs)\right]}\Bigg[\Phi\left(1,\frac{1}{q-1},\frac{1}{(q-1)\bs\omega}\right)\nn\\
&-&\frac{1}{(q-1)\bs\omega}\Phi\left(1,\frac{q}{q-1},\frac{1}{(q-1)\bs\omega}\right)\Bigg],
\eea
where
\bea
\text{Tr}_R\left[\hvrr^q_R(\bs)\right]=\left[(q-1)\bs\omega\right]^{\frac{q}{1-q}}\zeta_\text{H}\left(\frac{q}{q-1},\frac{1}{(q-1)\bs\omega}\right).\nn\\
\eea

\bibliography{bibNegSE.bib}
\end{document}